Presented at the 59th International Astronautical Congress, Glasgow, Scotland, (2008).
IAC-08-B2.5.2

# LEO-BASED OPTICAL/MICROWAVE TERRESTRIAL COMMUNICATIONS

Dr. Andrew Meulenberg
HiPi Consulting, New Market, MD, United States
mules333@gmail.com

Mr. Rahul Suresh
National Institute of Technology Karnataka, Surathkal, India
rsbullz@gmail.com

Mr. Shivram Ramanathan
National Institute of Technology Karnataka, Surathkal, India
rshiv.1987@gmail.com

## ABSTRACT:

*We propose an LEO-based communication system which is built by deploying circum-terra, optic fibers connecting hundreds of small (perhaps) phased-array-communications and RF-signal-transfer antennas around the earth on multiple rings. The proposed LEO-earth connection will be through microwave links (Ku or low-Ka band) and many of the ISL's through the optic-fiber rings. Inter-ring connectivity could be with either high-Ka band or optical (laser) links. The initial ring would serve to complement existing terrestrial fiber networks; but, the system would expand with additional rings into non-equatorial planes to provide global connectivity. The proposed system would make use of connectivity & broadcasting capability of satellite constellations as well as the high throughput point-to-point capability of optic-fiber systems. The advantages, options, and economics of the proposed LEO optic-fiber / microwave communication system over existing terrestrial- and space-communication systems (of similar functionality) as well as the future development of the system are dealt with in the paper.*

## 1. INTRODUCTION

### 1.1 Purpose of the proposed system

The communication system proposed in this paper is stage one of a multi-purpose low-earth-orbit-based system of rings circling the earth, the "LEO ARCHIPELAGO™".

Early establishment of a global communication network in LEO would generate revenue for the subsequent development of the system. This would be followed by the deployment of a new concept [1] of space elevator called "sling-on-a-ring™." This less-expensive mass transport into LEO would facilitate growth of solar power stations in high LEO, providing renewable and clean energy that man, so badly, needs. Expansion of the system beyond the LEO would assist his venture into inter-planetary and deep space.

This paper specifically addresses development of the LEO communication system(s).

### 1.2 About the proposed system

Two significant breakthroughs in international telecommunication were the launch of Communication Satellites into space and the introduction of under-sea optical-fiber communication. The advent of Geo-stationary communication satellites made it possible to cover large areas of the earth's surface with ease. They are ideal for point-to-multipoint communication. Satellite circuits are also terrain independent, hence making it possible to connect places where terrestrial networks cannot reach. But, the ~0.5 second round-trip time lag associated with GEO satellites makes it less-that-ideally suited for interactive-services[2]. On the other hand, terrestrial optic-fiber networks serve as high-throughput point-to-point communication, playing a pivotal role in satisfying our insatiable hunger for bandwidth.[3,4] But, it is still haunted by all the problems faced by terrestrial communication network (such as high deployment and repair costs involved with submarine cables and inaccessible terrain).[5]

Another problem with terrestrial optic-fiber networks is that it is the economic necessity to lay cables only for the more crowded urban regions leaving the rural areas (which form a large part of the earth) without high-speed communication capabilities.[6] Space communication systems have an advantage in this respect as they can achieve global coverage more easily. With GEO crowded, it is worth considering options to put new systems into other (non-geosynchronous) orbits. Satellite constellations in lower earth orbits (LEO) solve some of the above problems, as they have negligible round trip delays (relative to GEO satellites), while still covering the entire globe. There have been several such systems already proposed for LEO, like Iridium and Globalstar, for mobile telephony and Teledesic for broadband internet applications. Appendix A compares the proposed system with some of the other communication systems.

We propose a set of communication systems based on the deployment of circum-terra, optic fibers. With hundreds of small optically-interconnected, microwave-communications and signal-transfer antennas speeding around the earth, this system is the inverse of the standard cellular configuration. The LEO-earth connection will be through microwave phased-array (or small directable) antennas, while the optical rings would act as superior intra-ring Inter Satellite Links (ISLs). This system would provide benefits utilizing the advantages of a satellite-communication system as well as an optical-fiber system.

In this paper, section 2 deals with the experimental stage of the system which is pivotal in establishing the technical feasibility of the system. Section 3 discusses the salient features of the proposed system, while section 4 addresses the possible expansion of the system for global coverage and other capabilities. Possible applications of the system are discussed in section 5 and potential problems are addressed in section 6. Appendix B gives a back-of-the-envelope cost analysis of the system.

## 2. EXPERIMENTAL STAGE

The experimental stage of our communication system, prior to the deployment of complete optic-fiber/conductive ring would involve testing and development of the control and stability functions of the ring structures and of the ring-based solar-power and communication systems.

Multiple 20 km optical fibers coupled with a conductive fiber (e.g.; aluminum-coated Kevlar), along with several small phased-array, or cluster array, communications and microwave-power-transfer antennas would be launched into LEO. Solar power converters and ion-thrusters would provide the necessary basis for testing communications and station-keeping equipment and procedures for the long-string satellite.

The purpose of this initial “test-bed” is to establish ring-based capabilities for intra-ring, space-to-earth, space-to-space and earth-to-space communications; as well as for power transfer, maintenance and repair, and mass transfer within the local system. Efficacy of inter-satellite-links through optical fibers, microwave- and laser-signal transmission to the earth (uplink and downlink), and techniques for microwave-power beaming to the earth, etc. would be tested. The stability of the various fibers in the severe thermal-transient environment would also be studied.

This experimental stage would give a better understanding of the potential problems encountered (in terms of stability and power losses) and could hence play a crucial role in refining details prior to deployment of the full system.

Successful testing of the experimental stage would be followed by the initial operational phase of the communication system. This would be the LEO deployment of a near-equatorial, circum-terra, optic fiber (perhaps paired with, and connected to, a conductive fiber) along with the associated, sub-satellites (electronic and power "clusters" supporting the optically-coupled, phased-array-communications and RF-signal antennas).

## 3. SALIENT FEATURES

### 3.1 Circum-terra Optic Fiber in LEO

As the fiber lengths required for the optical rings are very long (around 40,000 km for the first ring at around 600 km and at 1200-2000 km for the final system), we resort to the use of single-mode optical fiber. The single-mode, perhaps hybrid, fibers (having lower attenuation and dispersion losses than their multi-mode counterparts) can run longer distances with the use of fewer repeater units. Single-mode optical fibers also have other advantages over multi-mode fibers in terms of long-haul transmission capability (e.g., higher bandwidth/throughput and freedom from inter-modal dispersion).

Due to the relatively fiber-friendly space environment, the layers of protective coating for the optical fibers can be drastically reduced as compared to undersea/terrestrial cables. This, and special handling, could permit the use of single-mode fiber with just a core and cladding, perhaps even without a single layer of sheath. To further reduce the mass-into-space required for the system, specialty fibers with reduced cladding (e.g. Corning RC-SMF) can be evaluated for handling and cost tradeoffs. Representative link budgets are presented in Appendix C. If the analyses in the appendices are realistic, then the system proposed would seem to have merit even without improvements that would inevitably result during the development. While it would cost more than the other LEO systems built and proposed, it has much greater capability in aspects important in today's world.

An example of options not considered at this point in the analysis would be the usage of independent "self-powered" repeater modules (each coupled with small solar modules). This can help power the repeaters and other such units in regions receiving sunlight directly (that is in a 90 minute cycle, a repeater is powered for approximately 45 minutes). Since there is maximum traffic during the daytime, this might prove advantageous. We can perhaps do away with need for the conductive ring and hence reduce the mass and launch costs of the optical rings by using this concept of "self-powered" repeaters.

### 3.2 Inter-satellite links: *within and between rings*

Inter-satellite links are of two types in our system: links between substations on the same ring (for whatever purpose) and links between different rings (crossing at slant angles – see section 4). Intra-ring ISL's are straightforward (for broadband communications along the ring path), thereby reducing the need of the complex routing and networking associated with inter-connectivity of multiple satellites in different orbits. Inter-ring information transfer would take place through antennas placed strategically along each ring for this purpose. The upper portion of the Ka Band (~60 GHz) is the choice for inter-ring ISL's since it is highly attenuated by the atmosphere and therefore cannot be used for space-to-ground applications and would not interfere with terrestrial applications at this frequency.

While microwave links are generally used in applications that require large-area coverage and long distance connectivity (e.g., point-to-multipoint, space-to-ground communications), they are band-limited to the order of few hundred Mbps. [7] This slow link in the communication system could reduce the bit rate of flow through the entire system to a few Mbps and also limit the scope for further improvement in the bit rate required for the future. However, with hundreds of transceivers on the optic rings, the individual link requirements are significantly reduced if the number of optical rings (or of ground stations) is also high. The limitations under these conditions are primarily those of switching, packaging, and directivity. Therefore, the whole system remains robust in terms of connectivity and extremely high effective throughput.

Optical ISLs, on the other hand, are ideal for high data-rate point-to-point applications due to the high focusing gain of laser beams. However, Free-Space Optical Communication (FSOC) links are not presently suitable for large distances because the narrow, high-gain, laser beams (required because of limitations imposed by available laser-diode powers – which problem may be reduced by development over time) must be pointed and maintained very precisely. This requires precision tracking telescopes with feedback loops. [8] The requirements common to small free-flying satellites, exacting station-keeping, greater platform rigidity, and/or reduced vibration, are reduced when the ring structure benefits are applied. There is even the problem of atmospheric attenuation and lower bit rates due to the optical beam spreading for long-range ISLs in LEO. These problems can be addressed by using the fiber-optic links as ISLs within the ring and then using the FSOC links over shorter distances between rings. This combination provides a high bit-rate/throughput communication system that can work efficiently without a requirement for precision tracking telescopes.

3.3 Microwave LEO-Earth connections

Use of higher Ku band/ lower Ka Band (15-40 GHz) is proposed for earth-to-space and space-to-earth signal transfer (because of the reduced atmospheric attenuation), especially for point-to-multipoint applications. Microwave communication is a well established and proven technology. It provides wide coverage; hence it is ideal for multicasting service, and omni-directional applications.[3] Also the microwave signals are not as adversely affected by atmosphere and clouds as optical signals. By offering faster throughput, dynamic bandwidth allocation, and smaller antennas than C or X band, Ka-band promises better satellite broadband service, more options and lower expenditure. [9] However, Ka band is badly affected by atmosphere and weather, nevertheless mitigation techniques such as adaptive resource sharing, adaptive coding and modulation, adaptive data rate switching and adaptive beam forming, etc. have been developed to compensate for high attenuation in Ka band.

Appendix A-2 compares the link budget for LEO-earth C Band, LEO-earth Ka Band and GEO-earth C Band communication systems.[9, 10] Situations are different now than when systems were being configured 20-40 years ago. Communications usage (e.g., internet, high-definition-TV, and cellphones) and technology advances (hardware and software) create new drivers for system choices today.

3.4 Optical LEO-earth connections

Use of optical frequencies for earth-to-space and space-to-earth signal transfer has been proposed, especially for very-high-throughput point–to-point communication. Optical free-space beaming to earth could be used to link the proposed space-based communication system with ground stations. Radio frequency can transmit data much farther than FSOC, but its bandwidth is limited to the order of a few hundred Mbps. Available FSO systems provide a bandwidth of up to 2.7 Gb/s. Links at 160 Gbps has been successfully tested in laboratories; speeds could potentially be able to reach into the Terabit range.[8] Also the narrow beam of the laser makes detection, interception and jamming very difficult, hence making it suitable for transfer of financial, legal, military, and other sensitive information.[8]

The main drawback of optical communication systems with ground stations is the influence of atmospheric effects in the form of optical scintillation and of attenuation by clouds.[11] On dense foggy days or cloudy days, attenuation due to atmospheric effects can exceed 200 dB/km; thereby making communication (even from LEO) almost impossible.[12] Adaptive optics (AO) can compensate for optical scintillation;[12, 13] but, they are probably not needed to LEO. To reduce the effect of attenuation by clouds, a site diversity technique must be used in order to increase the link availability to satellites. For the U.S.A. it has been shown that over 90 % availability can be achieved using 4 to 5 optical ground stations set up in the mainland and in Hawaii.[14]

Development of sophisticated digital signal processing techniques and algorithms (using tools like Lasercom network optimization tool) would enable our system to provide optical beaming to earth (provided large number of redundant interlinked ground stations are provided), on clear and less cloudy/foggy days. However, a better option might be the use of a high-altitude (5000 m) aerostat.[15] Use of such a tethered "blimp" as a "ground-station" (strategically located in low-cloud regions away from commercial airlanes) would allow a much-longer and more-reliable optical-communications pathway to LEO. It would only have high signal attenuation when actually enveloped by an occasional severe thunderstorm. (If the storm were even a few km away, multiple paths to the different sub-satellites on the LEO ring would permit broadband optical signals to be unaffected.)

### 3.5 Presence of unmanned robots:

One of the major disadvantages of sub-marine optical fibers is the difficulty, time, and finance involved with cable repairs. Cable breaks are by no means a thing of the past, with more than 50 repairs a year in the Atlantic alone,[5] and significant breaks in 2006 and 2008. It often takes several days to repair major disruptions. Traffic losses of 70 percent in Egypt and 60 percent in India were reported during the 2008 submarine cable break. Such incidents are economically and politically unacceptable in the present "communication-era".

Keeping these things in mind, we propose development and deployment of unmanned robots in space for the construction and maintenance of the ring system. These robots will be equipped with necessary splicing gear (mechanical splices and arc welding equipment) and supplies. These robots would help in rapid servicing (probably hours rather than days) for any cable break.

## 4. EXPANSION OF THE SYSTEM

Primary purpose of the initial equatorial circum-terra ring would be to link the N–S submarine cables and hence complement the existing terrestrial network. Successful deployment and functioning of this ring would generate revenues to support growth of the system.

The next proposed step in the development of the system would be expansion into non-equatorial orbits to attain global coverage. Inclined-orbit rings above fixed ground points (higher-LEO, earth-synchronous rings) would be sequentially deployed. Applying the Walker satellite-constellation model to the proposed communication ring system,[16] most of the areas between +70 to -70º latitudes would be covered (which is most of the human-inhabited areas). Specific rings, catering mainly to high-revenue, or to the earthquake prone, regions like Japan, or to highly-populated regions like India and China would be deployed first.

## 5. SPECIFIC APPLICATIONS

### 5.1 Broadband

#### 5.1.1 Bi-directional Broadband Applications

Being a LEO satellite constellation (at a >1000 km above the surface of the earth), the proposed communication system has lower latency (point-to-point round-trip of 10 - 160 milliseconds) as compared to that of geostationary satellites (always ~430 ms). This benefits some broadband applications like broadband Internet, rural resource centers (for tele-education and tele-medicine in rural areas), videoconferencing, and other interactive services[2]. The distributed or decentralized topology of the LEO system has higher reliability[2] and the smaller footprint of LEO satellites enable frequency reuse and hence results in more efficient use of limited spectrum resources. The large number of small sub-satellites also reduces unit design cost involved in manufacturing and creates a system with very large total capacity and redundancy, which allows a lower cost of service. Thus, the system can provide affordable broadband services for developing and under-developed countries.

#### 5.1.2 Interlinking with terrestrial broadband

The high bandwidth offered by the optical rings can be used to complement the existing submarine fiber networks and can also aid in linking terrestrial fiber networks like (e.g., Eastern Africa Submarine Cable System and Asia-Pacific Cable Network). Initially, the system would feed the North-South running fibers. The gradual development of this system would eventually result in easy access of broadband services and utilities at even the most remote corners of the world.

5.2. Mobile Telephony

Satellite constellations in general (and this system is no exception!) can be used to provide mobile telephony and low-bit-rate communication services. These have already been tested and established in the Iridium and Globalstar systems. [17] These systems must interoperate within Global System for Mobile communications (GSM) and with other terrestrial cellular networks, and generally offer basic fax, low-speed data and paging facilities (2400/9600 bps) alongside voice services.[18] This can be done easily as these mobile satellite systems use the inverse of the standard cellular configuration (moving cells and relatively static or slow-moving users). The system, on further expansion into slant orbits, higher orbits and eventually into MEO is capable of being a standalone space-based global-communication system.

5.3. Messaging

The proposed communication system can provide low-bandwidth send, receive and broadcast services for a variety of applications, including paging, tracking, and remote data gathering on the lines of already existing systems like Orbcomm.

5.4. Television broadcasting

Once the communication system is expanded beyond the equator and complete coverage to particular regions is achieved (ie every point in the identified region comes under the footprint of at least one satellite at any given point of time), simple point to multipoint connectivity feature of each satellite is used to transmit to all receiving stations (or homes in case of DTH) within its footprint. With the ISL's being fairly simple, and with further advances in networking, continuous switching of transmitting stations along with buffering of data would enable this technology .

5.5. Direct-to-home (DTH) TV systems

The main requirement of DTH digital high-definition TV (HDTV) is very-high bit rate and hence band width. The techniques of digital video compression and digital modulation technology have made it possible today to have fairly acceptable DTH reception at 110-120 Mbps in the Ku band.[19, 20] Since bandwidth in the proposed communication system is not an issue (Ka/Ku band for space to earth transmissions and optical bands for ISL's provides very-high band widths) and global coverage is planned, DTH HDTV broadcast would be a major profit maker for the system. Smaller receive-antenna arrays could be used for medium-bandwidth customers (e.g., those requiring only basic service – with fewer, but higher-power channels). These services could be controlled to provide specific regions with particular language broadcasts.

5.6. Broadband mobile

Since large antennas are not required for point-to-point broadband or broadcast in the LEO system, mobile systems (buses, trains, ships, and planes), would avail themselves of this service.

5.7. Disaster management

Being at low altitude the LEO communication system can also prove useful in disaster management by sensing the environment for changes indicating the arrival of natural disasters like storms, tsunamis, etc and thus be instrumental in saving millions of lives. Since a small auxiliary unit is the only thing required for establishment of local (albeit short-range) cell-phone connectivity, airdrops into an emergency situation could immediately restore communications when all else has failed.

## 6. POTENTIAL PROBLEMS

6.1 Legal and regulatory issues:

Since the magnitude of proposed communication system is enormous and is ultimately intended to provide connectivity to each and every corner of the world, it is obvious that several regulatory and legal hurdles have to be overcome before the system actually becomes functional.

The first regulatory channel / requirement would be to obtain all desired frequencies in the spectrum allocation. The proposed system intends to use 4 distinct frequency ranges in the electromagnetic spectrum- Ku band for mobile communication, lower Ka band (30 GHz) for broadband applications, upper Ka band (60GHz) for inter-ring satellite communications and optical frequencies (along optic fiber rings and perhaps point to point secure beaming to earth). Possible hurdles in obtaining desired frequencies include unavailability of bandwidths in the desired frequency range and opposition from radio astronomers and other users, or potential users. The most amiable solution will have to be explored so that desired frequencies are obtained with minimum challenge. (Guarantees of optical and radio telescopes in space might be a condition that could be negotiated – assuming a low-cost mass transfer system to LEO is available.)

The second issue would involve international co-operation for space assets (orbits). As the initial ring system passes over several countries and eventually the system is intended to provide global coverage, agreements will have to be made with all involved countries. The best solution would be
to make all the involved countries stake-holders of the project rather than just beneficiaries. The first optic-fiber ring would be along the equatorial plane. Most of the countries in this region are developing and do not have a well-established communication network; hence it is believed that most of them would be willing to show enthusiastic participation in the project. More international agreements will have to be signed as the system expands to non-equatorial orbits. (However, it is likely that these required international agreements will need to be in place before system hardware is even begun.)

Last, but not the least, once the system is fully deployed (with equatorial and non-equatorial circum-terra optic-fiber rings and hundreds of mini broadcast stations), those orbits would be closed for all other future space endeavors. This would naturally attract opposition from several lobbies. Also a broken sling could prove dangerous to the orbits about it. Since our slant-orbit constellation would begin at 1200 km altitude and the space above is relatively free (due to the inner Van Allen radiation belt), the broken-fiber problem may not be a great concern from that perspective.

The legal and regulatory issues may be challenging, but not impossible to overcome. This fact was successfully demonstrated by Iridium[16] when it defeated an attempt by radio astronomers to prevent it from getting the 5.15 MHz bandwidth in the L band. Iridium also successfully obtained approval from key countries where gateways were strategically placed.

Another example of this international agreement was that of the Teledesic system acquiring rights to use a part of the Ka band. After many appeals and discussions, the FCC granted 400 MHz of spectrum at 28.7 to 29.1 GHz (uplink) and at 18.9-19.3 GHz (downlink) for this system.[21] Even though the Teledesic system was never implemented, this in itself was a major achievement for the system. These frequencies being allotted for NGSO (Non-Geostationary-Satellite-orbit) systems might still be available as the Teledesic system never went into orbit.

## 6.2 Other issues

### 6.2.1 Space debris

The ubiquitous space debris problem has an effect on the proposed system as well. The space debris in LEO orbits is dangerous to the fibers we are putting up. This effect might be somewhat reduced by the planned reduction in the cladding and/or sheath and hence the interaction cross section of the fibers. While this might reduce the probability of a space debris hitting and damaging the fiber, it also removes some protection that could reduce the damage when collision does occur. The larger space debris particles are capable of damaging the mini-satellites also.

### 6.2.2 Radiation

The proposed system will extend into the inner Van Allen Belt. Since, the electronic components are distributed among many identical sub-satellites, any required radiation shielding would be costly in terms of mass in orbit. The sub-satellites are small and bound to a ring. Therefore, there will be little self shielding from internal components, subsystems, and structural panels normal to larger free-flying satellites. However, this system is part of a larger system that would utilize large area thin-film arrays to first reduce space debris in LEO and then be moved into polar orbit to eliminate the trapped radiation belts. Periodic influx of radiation from solar flares would be rapidly absorbed/scattered, so that total dose effects over extended periods would be minimized.

#### 6.2.3 Temperature

The temperatures in space can be extreme (about -140°C in the night side and 80°C in the day side). This may have an effect (physical or optical) on the optical fiber used in the proposed system, as normal fibers are specified for operation in the range -55°C to +85°C. The low temperatures may cause them to become brittle and micrometeoroid or space-debris impact could have a much greater impact. Thermal mismatches must also be eliminated over a larger temperature range than presently specified.

## 7. CONCLUSION

The proposed global communication system is an idea that has immense potential to reap the technological benefits of both terrestrial and space communication systems. Initially as a supplement to existing communication systems and eventually as a stand-alone system, it has the capability to satisfy the ever-growing need for communications all over the world. Several potential show-stoppers that could deter the system's growth will have to be addressed through further research. Nevertheless, changes in both communications technology and usage as well as potential changes in launch costs and space-systems development could greatly alter the trade space for evaluating space communication systems.

## APPENDICES

### Appendix A1: Comparison of proposed system with other LEO communication systems*

| | Iridium | Teledesic | Proposed system |
|---|---|---|---|
| Orbit type | LEO | LEO | LEO |
| Altitude (kms) | 780 | 695-705 | 1200 kms |
| Period (min) | 100.1 | 98.8 | 109 |
| Number of orbital planes | 6 | 21 | 16 |
| Total Number of satellites | 66 | 840 | 500-600 |
| Footprint diameter (km) | 4700 | 1412 | 1200 |
| Uplink and downlink frequencies | L and Ka band | Ka band | Ku and Ka band |
| ISL frequencies | 22.5 GHz | 60 GHz | 60 GHz, optical |
| Satellite mass (kg) | 771 | 700 | ? |
| Bit rate | 2.4 kbps | 16-2048 kbps | *similar to teledesic*† |
| Coverage | Global | Nearly global | Eventually global |
| Broadcast services | No | No | Yes |
| Broadband fixed services | No | Yes | Yes |
| Broadband mobile services | No | Yes | Yes |
| System costs (est.) | $5.7B | $10B | $10 - 20B |

* Numbers vary from source to source for many of the items [15]

† Uncertain, or to be determined

Appendix A2: Comparison with individual GEO and terrestrial networks*

| Environment | GEO<br>36000km orbit | Fiber-optic deployment<br>Terrestrial and under-water | Proposed system<br>Orbit at 1200 km |
|---|---|---|---|
| Installation/Deployment | only 3 launches | difficult in harsh environments | *few hundred launches |
| Coverage | Global using 3 satellites | Terrain dependent | Global using 600 minisatellites |
| Frequencies used | C, Ku and Ka | Optical | Ku, Ka and optical |
| Latency | 1/2 sec | Negligible | Negligible |
| Broadcast services | Yes | No† | Yes |
| Interactive services | No† | Yes | Yes |
| Broadband fixed services | Yes | Yes | Yes |
| Broadband mobile svc | Yes | No† | Perhaps yes |
| Remote sensing | No | No | Yes |
| Disaster Management | No | No | Yes |
| In case of major failure | Severely affected | Very severely affected | *Not badly affected |

Appendix B: Initial System Cost Analysis:

Assumptions

1. Number of sub-satellites on the initial ring is up to 40
2. Fiber cost is around $15 per km.
3. Costs of repeater units, dispersion compensation units, transceivers, etc put together will be almost twice the cost of fiber.[22]
4. Mass density of fiber is around 140 g/km.
5. Cost of each solar power module is $1 million
6. Launch costs to LEO ~ $2000 per pound considering non-western launch vehicle.[23] Launch costs can be as low as $1400 per satellite through private launches through Lockheed Martin. This can bring down our costs by another $25 million.
7. The Teledesic system had a target of $5.5 million per satellite which is substantially less than $100 million per satellite which is the usual cost of a communication satellite.[23]

| | |
|---|---|
| Fiber | 10 tonnes |
| Satellites (800kg per sat) | 16 tonnes |
| Other or overhead | 8 tonnes |
| Total | 34 tonnes |

Table B1. Rough estimate for mass of the initial communication system

| Item | Cost |
|---|---|
| Fiber costs | $2 million |
| Satellites w Microwave antennas, Telescopes and lasers, onboard electronics, etc (40 satellites at $10M each) | $400 million |
| Launch costs | $200 million |
| Assembly and ground control | $20 million |
| Solar power modules | $40 million |
| on-orbit install-and-repair "robots and "tugs" | $200 million |
| Experimental communication and power links | $100 million |
| Development costs* | $200 million |
| International agreements, Freq allocation costs, etc | $100 million |
| Contingency (e.g., pay for shuttle "service" call) | $100 million |
| Total* | ~$1.4 billion |

Table B2. Rough estimate for costs of the initial communication system. The costs estimated are for the initial ring only. Subsequent rings would have fewer sub-satellites (higher altitude => larger "foot print" for each sub-satellite) and substantially lower costs.

Appendix C: Fiber-optic link budget

| | |
|---|---|
| Connectors(2*0.75 dB/connector) | 1.5 dB |
| Fusion Splices (8*0.02 dB/splice) | 0.16 dB |
| Fiber dispersion | 1 dB |
| SPM Margin (Self phase Modulation) | 0.5 dB |
| XPM Margin(Cross phase Modulation) | 0.5 dB |
| DCU Compensation(Dispersion compensation unit) | 6 dB |
| FWM(Four-wave mixing) | 0.5 dB |
| SRS/SBS (Stimulated Raman Scattering) | 0.5 dB |
| PDL(Polarization dependent loss) | 0.3 dB |
| PMD (Polarization mode dispersion) | 0.5 dB |
| Amplifier gain tilt (due to non-flat gain spectra) | 3 dB |
| Receiver sensitivity tilt (wavelength dependence of PMD) | 0.5 dB |
| Transmitter chirp | 0.5 dB |
| AWG cross-talk | 0.2 dB |
| Optical safety and repair margin | 3 dB |
| **Total Loss (dB)** | **18.7** |

Table C-1: Fiber-optic Long-Line Component Losses[24] per span

Un-repeatered length of fiber and number of repeaters required in a circum-terra ring.

We assume operation at a bit rate of 10Gbps, for which the corresponding receiver sensitivities for an avalanche photodiode (APD) are around -30dBm (max). Fiber losses are assumed to be 0.2dB/km. Temperature sensitivities have not been included in our rough calculations.

| | |
|---|---|
| Transmitted power | 0 dBm |
| Receiver sensitivity | - 30 dBm |
| Power budget | 30 dBm |

Table C-2: Optical power budget:[24]

From table C-1, span losses (w 3dB margin) = 18.7 dB
Allowed span loss = budget – loss =30-18.7dB = **11.3dB**
Length of each span is (30-18.7)/0.2 = 56.7 km
Hence, required no. of repeater units in a ring = **76**

| LINK BUDGET | UNITS | LEO C BAND | LEO Ka BAND | GEO C BAND |
|---|---|---|---|---|
| Altitude | kms | 500 | 500 | 36000 |
| *Transmit power* | *dBW* | *17* | *17* | *17* |
| Frequency | GHz | 5 | 30 | 5 |
| Wavelength | mm | 60 | 12 | 60 |
| Tx antenna diameter | m | 0.18 | 0.04 | 7 |
| Tx antenna gain | dBi | 18.7 | 18.7 | 50 |
| Feeder loss | dB | 3 | 2 | 3 |
| EIRP | dBW | 32.71 | 33.71 | 64.19 |
| Pointing loss | dB | 1 | 1 | 1 |
| Polarisation loss | dB | 0.5 | 0.5 | 0.5 |
| Beam divergence | degree | 22.5 | 22.5 | 0.6 |
| Path loss (Free space loss) | dB | 160 | 174 | 198 |
| Rx antenna diameter | m | 0.037 | 0.037 | 0.037 |
| Rx antenna gain | dBi | 4.7 | 18.7 | 4.7 |
| Feeder loss | dB | 3 | 2 | 3 |
| Atmospheric losses | dB (max) | 2 | 10 | 2 |
| Rain losses | dB (max) | 0.2 | 10 | 0.2 |
| *Receive power* | *dbW* | *-126.22* | *-142.9* | *-131.8* |

Table C-3: Uplink/Downlink link-budget (Ka relative to C-band).

Losses in a Ka-band link budget, relative to C-band, from rain is the primary deterrent to use of this frequency for satellite communications. However, acceptability of occasional outages or use of auxiliary units with somewhat larger antennas could make this a frequency of choice for many countries and applications.

[6] A. Kupfer, "Craig McCaw Sees an Internet In the Sky," *Fortune*, 27 May '96 referenced http://www.smgaels.org/physics/97/CDIAZ.HTM

[7] Press release "CableFree launches range of High Performance Broadband Radios," http://www.cablefreesolutions.com/pdf/PR%20Broadband%20Radio%20Launch%20-%20Final.pdf

[8] W. Morgan, "Inter-satellite links," Space Business International, Quarter 1 1999

[9] "Market Dynamics Of Satellite-Delivered Ka-Band Broadband Service 2003," Frost and Sullivan, http://www.satelliteonthenet.co.uk/white/frost1.html

[10] J. P. Silver, "Satellite Communications Tutorial," online http://www.odyseus.nildram.co.uk/Systems_And_Devices_Files/Sat_Comms.pdf

[11] L. C. Andrews and R. L. Phillips, Laser Beam Propagation through Random Media, SPIE, Optical Engineering Press, Bellingham, Wash., 1998

[12] F. Levander and P. Sakari, "Design and Analysis of an All optical Free-space Communication link," University essay from Linköpings universitet / Institutionen för teknik och naturvetenskap 05/23/2002 www.divaportal.org/diva/getDocument?urn_nbn_se_liu_diva-1198-1__fulltext.pdf

[13] R. K. Tyson, "Adaptive optics and ground-to-space laser communications," Appl. Opt. 35, 3640-3646 1996.

[14] R. Link, R. Alliss, M. E. Craddock, "Mitigating the impact of clouds on optical communications," Proc. SPIE 5338, 223-232 (2004).

[15] E.g., TCOM's 71M® Aerostat http://www.tcomlp.com/71M.htm

[16] O. de Weck, R. de Neufville, D. Chang, M. Chaize, "MIT INDUSTRY SYSTEM STUDY: Communication systems constellations, Unit 1 "Technical Success and Economic Failure" web.mit.edu/deweck/www/research_files/comsats_2004_001_v10/Unit1%20Success%20and%20Failure/Unit1_lecture.ppt

[17] L. Golovanevsky, G. Pattabiraman, K. Thakar, R. Yallapragada, "Globalstar Link Verification"- Proceedings of Wireless Communications and Networking Conference, 1999 WCNC. IEEE, 1182-1187 vol.3

[18] L. Wood, "Internetworking with satellite constellations" PhD Thesis, University of Surrey, June 2001

[19] D. C. Agarwal, Satellite communications, revised by A.K. Maini, Delhi : Khanna Publishers, 1998.

[20] K.C.Bhasin, "Direct-To-Home TV: Transmission And Reception," http://www.electronicsforu.com/EFYLinux/efyhome/cover/aug2003/DTH-TV.pdf

[21] Authorization letter by FCC http://www.fcc.gov/Bureaus/International/Orders/1997/da970527.txt

[22] M. Fullenbaum, "Next Generation of Transoceanic Submarine Solutions," PTC'06 Proceedings

[23] Futron Corporation, "Space Transportation Costs: Trends in Price Per Pound to Orbit 1990-2000," (White paper) September 6, 2002

[24] T. Antony and A. Gumaste, "DWDM Network Designs and Engineering Solutions" Cisco Press, Series: Networking Technology, Pub. 12/13/02 http://www.ciscopress.com/articles/article.asp?p=30886&seqNum=6